\documentclass[journal,transmag]{IEEEtran}
\usepackage{graphicx}

\ifCLASSINFOpdf
\else
\fi
\begin{document}
%
\title{Bufferless transmission in complex  networks}


\author{\IEEEauthorblockN{Cunlai Pu\IEEEauthorrefmark{1},
Wei Cui\IEEEauthorrefmark{2},
Jiexin Wu\IEEEauthorrefmark{1},
Jian Yang\IEEEauthorrefmark{1},~\IEEEmembership{Fellow,~IEEE}}
\IEEEauthorblockA{\IEEEauthorrefmark{1}School of Computer Science and Engineering, Nanjing University of Science and Technology, Nanjing 210094, China}
\IEEEauthorblockA{\IEEEauthorrefmark{2}EMC corporation, Beijing 100027, China}
\thanks{
Corresponding author: Cunlai Pu (email: pucunlai@njust.edu.cn).}}

%



\IEEEtitleabstractindextext{%
\begin{abstract}
Complex bufferless networks such as on-chip networks and optical burst switching networks haven't been paid enough attention in network science. In complex bufferless networks, the store and forward mechanism is not applicable, since the network nodes are not allowed to buffer data packets. In this paper, we study the data transmission process in complex bufferless networks from the perspective of network science. Specifically, we use the Price model to generate the underlying network topological structures. We propose a  delivery queue based deflection mechanism, which accompanies the efficient routing protocol, to transmit data packets in bufferless networks. We investigate the average deflection times, packets loss rate, average arrival time, and how the network topological structure and some other factors affect these transmission performances. Our work provides some clues for the architecture and routing design of bufferless networks.
\end{abstract}

\begin{IEEEkeywords}
Bufferless networks, Deflection mechanism, Routing protocol.
\end{IEEEkeywords}}

\maketitle

\IEEEdisplaynontitleabstractindextext

%
\IEEEpeerreviewmaketitle

\section{Introduction}
%
%
%
%
\IEEEPARstart{N}{owadays}, we build a lot of communication networks according to our needs with advanced technologies. Based on the buffer availability at network nodes, these communication networks can be divided into buffer-based communication networks and bufferless communication networks\cite{schwartz95}.The former includes the Internet, mobile communication networks, sensor networks, mobile ad hoc networks, and so on. The latter includes the on-chip communication networks \cite{demicheli17,jantsch03}, optical burst switching network \cite{chen04,robertazzi}, etc.
Information transmission is one of the fundamental functions of all these complex communication networks. Different types of communication networks have their own information transmission mechanism. For the buffer-based communication networks, they use the store and forward mechanism \cite{fultz71,baran64}. When the traffic flow is large, nodes can not manage all their traffic load due to limited processing capacity, and thus some data packets will be stored in node buffers for subsequent forwarding.
This kind of store and forward mechanism needs lots of network resources and consumes plenty of node energy to support large traffic and high reliable data communication, more suitable for large communication infrastructures. The bufferless communication networks do not adopt the store and forward mechanism, since they have energy and resource constraints and real-time transmission requirement. Instead, they usually use static routing protocols combined with random deflection mechanism for data transmission \cite{acampora92,haeri16}. In this case, data packets are transmitted with a given static routing protocol with priority. This will cause the packets compete for the port forwarding, since the node port forwarding capacity is limited. Failed packets are forwarded from other free ports according to the given random deflection mechanism. If all node ports are busy, the failed packets are discarded.

Network science developed rapidly in the past decade, and relevant theories have been applied to the communication networks \cite{bal16}. However, most work is about the buffer-based complex networks. Researchers were dedicated to improving the network transmission capacity of buffer-based complex networks based on novel strategies \cite{schen12,yanghx08,pucl12,yangx17,Pucl13},  which can be basically classified into three categories. The first category is about optimizing the network topologies.
Liu et al. \cite{zliu} pointed out that removing some links between the core nodes can effectively make the data flow avoid the core nodes, thereby reducing traffic congestion. Zhang et al. \cite{gqzhang} found that deleting some links between large-betweenness nodes can also reduce the traffic congestion. Huang et al. \cite{whuang} further improved the above two kinds of link optimization strategies by defining a node-difference index. Later, they further improved the network capacity by adding some links between distant nodes and between the neighbors of large degree nodes \cite{Huangw}.

The second category is to optimize the network resources. Kim et al. \cite{kim08} studied the relationship between the delivery capability and the load of nodes in different networks. They found that the node's delivery capability in the real systems is not fully utilized due to the fluctuation of the network data traffic. This problem is more serious among nodes of small delivery capability.
Liu et al. \cite{liuz06} assumed that the node's capabilities of generating and delivering packets are related to the node degree, and further studied their impact on traffic congestion.  Gong et al. \cite{gongx08} found that there is a maximum network capacity when the network structure and routing strategy are given, which is inversely proportional to the average transmission path length. Ling et al. \cite{xingl13} considered both the node's delivery capability and link bandwidth constraints, and they found that allocating the resources based on the algorithmic
 betweenness of nodes can achieve the maximum network capacity. Liu et al. \cite{liuh15} considered the heterogeneous data flow generated by the nodes, and optimized the network flow and the node delivery capability to maximize the network utility.

The optimization of network structure or network resource is related to network hardware, and thus it has high cost and sometimes it is not feasible in real applications.
The third category is the optimization of network routing protocols. This kind of work usually involves only the network software, which is relatively simple to implement and the cost is relatively small. For example, Yan et al. \cite{gyang046} proposed an efficient routing algorithm based on the degrees of on-path nodes  with a control parameter.  Du et al. \cite{duwb13} further considered the forwarding priority of the packet based on the global routing algorithm proposed by Yan. Wang et al. \cite{wangwx06} proposed a random forwarding strategy based on neighbor node degree. They further considered the load information and proposed a local dynamic routing strategy \cite{wangwxaa}.  Ling et al. \cite{ling2010} proposed a global dynamic routing protocol which selects the optimal path of the smallest path load. Wu et al. \cite{zxwu08} proposed a routing strategy based on both the neighbor node degree and the shortest path length information. Yang et al. \cite{yanghx14} proposed an adaptive routing strategy based on node distance and node load. The above-mentioned routing algorithms can effectively improve the network capacity, but these algorithms are specifically for single-layer complex networks. Recently, the routing algorithms for complex multi-layer networks have attracted more attention \cite{tan14,du72,puc1626,zhouj13}.

The work described above is about buffer-based communication networks. The theories and methods from network science area should also be applied to bufferless communication networks. Different from the buffer-based communication networks, bufferless communication networks are more concerned with data loss rate and average deflection times of packet than    network capacity \cite{michelogiannakis,gnychis}.  From the angle of network science, we study the data transmission process in bufferless communication networks. Specifically, we dedicate to applying complex network models and routing strategies proposed for general complex networks to bufferless communication networks. We propose a new random deflection strategy to deal with the forwarding competition problem in bufferless communication networks. Through simulation, we investigate how  various factors  influence transmission performances.
\section{The network model}
We use Price model \cite{newman2010} to generate the underlying network topological structures. Price model is a typical model to generate scale-free networks. Compared to BA model \cite{barabasial99}, Price model can adjust the power-law parameter, thus adapting the node degree distribution. The steps of  fast algorithms of Price model are given as follows:

(\romannumeral1)  Initially, we generate a fully connected directed graph of $m_0$ nodes. Then, we put the labels of nodes that each link points to into an  array  $Arr$.

(\romannumeral2) We set a parameter $P\in [0,1]$. For $t=1, 2, \cdots, N-m_0$, we take the following actions:
(1) We generate a random number $r\in [0,1)$;
(2) If $r<P$, we randomly select an item from $Arr$;
(3) if $r\geq P$, we randomly select a node;
(4) We do step (1)-(3) $m$  times (maybe more times to avoid repeated choices), and  get $m$ links of the new node added at time $t$. We put the labels of the $m$ selected nodes into $Arr$. Note that for a new node, the $m$ links should point to $m$ different nodes.

Finally, we obtain an undirected scale-free network with $N$ nodes by  ignoring the directions of all links. The average node degree is $2m$. The power-law parameter is $\gamma=\frac{1+P}{P}$. If we set $P=0.5$, then $\gamma=3$, and Price model degenerates into BA model.
\section{Data transmission model and performances}
In buffer-based communication networks, non-sent  packets will be stored in the buffers at the nodes. In bufferless communication networks, all data packets should be forwarded at each time step. Packets are not allowed to waiting at the nodes. One the other hand, there is a delivery queue at each node. The delivery queue is used to store the  packets, which will be sent at the current time step. At the end of each time step, the delivery queue is empty.
\subsection{Transmission model}
 We study the data transmission process in bufferless networks. In our model, each node generates the data packets with  rate $\rho$. For example, $\rho=1.2$ means that each node generates one packet with probability 1 and another one with probability 0.2 at each time step. The destination of each packet is randomly chosen among all the nodes except the source node. The size of the delivery queue is equal to the delivery capability of node. For simplification purpose, we assume that node $i$ can at most deliver $C\ast k(i)$ packets at each time step, where $C$ is the node delivery coefficient  and $k(i)$ is the degree of node $i$. Note that the queue is for storing the received packets and the generated packets. When the queue is full, the node stops generating packets. If the current node is the destination of the packet,  the packet will be discarded from the queue immediately. We apply the efficient routing protocol \cite{gyang046} to our data transmission model. The basic idea of the efficient routing protocol is as follows: a path from node $a$ to node $b$ can be denoted as $P(a\rightarrow b):=a\equiv v_0,v_1,\cdots,v_{n-1},v_n\equiv b$, where $v_i$ is the $i+1$th node in the path. The transmission cost of this path is defined as follows \cite{gyang046}:
\begin{equation}
H(P(a\rightarrow b):\alpha)=\sum_{i=0}^{n-1}k(v_i)^{\alpha},
\end{equation}
where $\alpha$ is a tunable parameter. For all the paths from $a$ to $b$, the one with the smallest transmission cost is chosen as the optimal transmission path. If there is more than one optimal path, we randomly choose one as the transmission path. Obviously, the optimal transmission path is determined by the control parameter $\alpha$. When $\alpha=0$, the efficient routing protocol degenerates into the shortest path protocol.

For bufferless data transmission, we usually need certain deflection mechanism to deal with the forwarding competition. Failed packets will be delivered from other node ports  instead of the ports  determined by the static routing protocol. If there are no free port,  the failed packets will be discarded. Here, we propose a delivery queue based deflection mechanism. Specifically, a packet will check the delivery queue of the next-hop node given by the efficient routing before being sent. If the queue is full,  the packet will randomly select another neighbor node of the current node and check the queue status of the neighbor node. If the queue is also full, the packet will do another selection until finding a not-full neighbor node. If all neighbors' queue is full, the packet will be discarded.  This deflection mechanism accompany the efficient routing protocol to deliver the data packets in our model.
\subsection{Transmission performances}
For buffer-based communication networks, the network capacity is the first concern,  which measures the maximum allowed data packets generation rate of nodes under free flow state. However, for bufferless communication networks, packet loss rate and deflection times are more important than network capacity. Here we define the packet loss rate to be the ratio between the number of discarded packets (The arrival packets are not included.) and the number of generated packets for a large time period $T$, which is given as follows:
\begin{equation}
\eta(T)=\frac{n_l(T)}{n_g(T)},
\end{equation}
Where $n_l(T)$ is the number of discarded packets except the arrival packets during time period $T$, and $n_g(T)$ is the number of generated packets during time period $T$. For bufferless communication networks, the failed packets of the forwarding competition will be deflected to the alternative output node ports in the transmission. In our model, the failed packets will deflect according to the delivery queue based deflection mechanism. We define the average deflection times to be the ratio between the total number of deflection times of all packets  and the total  number of generated packets for a large time period $T$, which is as follows:
\begin{equation}
\omega(T)=\frac{n_d(T)}{n_g(T)},
\end{equation}
Where $n_d(T)$ is the total deflection times of all nodes  during time period $T$.  Also, we measure the average arrival time of packets $T_a$, which is the average number of transmission steps for a packet from the source to the destination. Through these three metrics, we can evaluate the transmission model, the performance of the routing algorithms,  and how the network topological structure affects the data transmission.

\section{Simulation results}
In this part, we investigate the transmission performances of our bufferless transmission model by simulation. The aim is to demonstrate how various factors affect the transmission performances. In the simulation experiments, we vary one particular parameter by keeping all the other parameters constant. Also, we let our model run a long time (T=1000)  in order to obtain a stable traffic flow.
\begin{figure*}[!t]
\centering
\includegraphics[width=5in,height=4.25in]{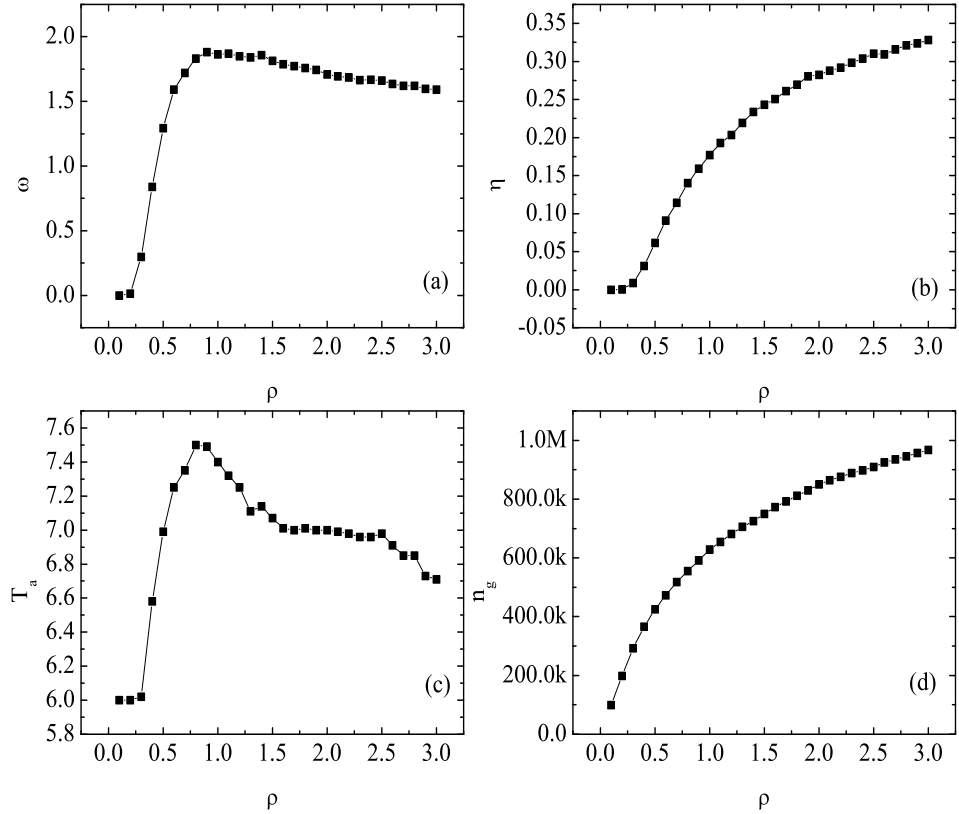}
\caption{Packet generation rate $\rho$ vs. (a) average deflection times $\omega$, (b) packet loss rate $\eta$, (c) average arrival time $T_a$, and (d) number of generated packets $n_g$. The other parameters are: network size $N=1000$, average node degree $\langle k \rangle=4$, power-law parameter $\gamma=3$, node delivery coefficient  $C=2$, and routing control parameter $\alpha=1$.   Each data point is the average of 100 independent runs. }
\end{figure*}

First, we study how the packet generation rate $\rho$ affects the transmission performances. In the experiments, we set network size $N=1000$, average node degree $\langle k \rangle=4$, power-law parameter $\gamma=3$, node delivery coefficient  $C=2$, and routing control parameter $\alpha=1$. The simulation results are present in Fig. 1, where each data point is the average of 100 independent runs. In Fig 1(a), we see that when the packet generation rate $\rho$ is very small, the average deflection times $\omega$ is zero. After $\rho$ exceeds the critical value 0.2, $\omega$ goes up abruptly and then reaches the peak,  and then goes down a little bit,  and finally converges. When the packet generation rate is very small, there is no packet forwarding competition, and all the packets are transmitted with the efficient routing protocol to the destinations. Thus, the average deflection times are zero. When the packet generation rate exceeds the critical value, the packet forwarding competition appears and becomes more serious with the increase of the packet generation rate. More and more packets are deflected to the non-full neighbors of current nodes. Thus, the average deflection times increase. 	When the packet generation rate is large enough, there may not be any non-full neighbors of the current nodes,  thus some packets are discarded directly instead of been deflected and the average deflection times decrease.  In Fig. 1(b), when the packet generation rate $\rho$ is very small, the packet loss rate $\eta$ is zero. After $\rho$ exceeds the critical value 0.3, $\eta$ goes up abruptly, and finally converges with the increase of $\rho$. When the packet generation rate is small enough,  there is no or slight forwarding competition. The packet transmission can be well managed by the efficient routing protocol and the  delivery queue based deflection mechanism. In this case, there is no packet loss. When the packet generation rate exceeds the critical value, the situation that all neighbors of the current nodes are full appears, and this leads to the packet loss. This problem becomes more serious with the increase of the packet generation rate.  In Fig. 1(c), we see that  the average arrival time $T_a$ goes up and then goes down with the increase of the packet generation rate $\rho$. This is because the increase of the average deflection times generally results in the increase of the average arrival time. However, when the packet generation rate is too large, many packets are discarded, the arrival packets are those with a few or no deflections, thus the average arrival time decreases.
 According to our model, when the delivery queue is full, the nodes stop generating packets, thus the number of packets inserted into the network converges  with the increase of packet generation rate, as shown in Fig. 1(d). This is why the average deflection times, the loss rate and the average arrival time converge finally.
\begin{figure*}[!t]
\centering
\includegraphics[width=5in,height=1.8in]{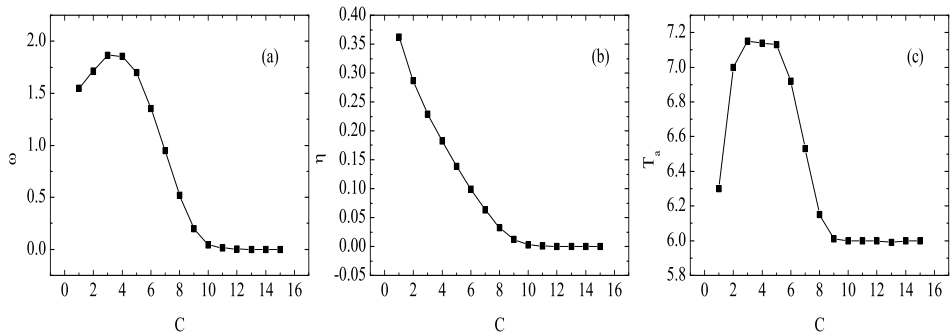}
\caption{Node delivery coefficient $C$ vs. (a) average deflection times $\omega$, (b) packet loss rate $\eta$, and (c) average arrival time $T_a$. The other parameters are: network size $N=1000$, average node degree $\langle k \rangle=4$, power-law parameter $\gamma=3$, packet generation rate  $\rho=2$, and routing control parameter $\alpha=1$.   Each data point is the average of 100 independent runs. }
\end{figure*}

Then, we investigate how the node delivery capability affects the transmission performances. Note that we defined the node delivery capability to be proportional to the node degree in the above. In the experiments, we vary the node delivery coefficient to change the node delivery capability. The relevant parameters are as follows: network size $N=1000$, average node degree $\langle k \rangle =4$, power-law parameter $\gamma=3$, packet generation rate $\rho=2$, and routing control parameter $\alpha=1$. The simulation results are shown in Fig. 2, where each data point is the average of 100 independent runs. In Fig. 2(a), we see that with the increase of the node delivery coefficient $C$, the average deflection times $\omega$ goes up first, and then reaches the peak, and then goes down and finally becomes zero. On one hand, the increase of node delivery capability increases the capability of receiving and forwarding the deflected packets, which results in the increase of average deflection times. On the other hand, the increase of node delivery capability decreases the number of packets needed to be deflected. Therefore, the average deflection times decrease until becoming zero.  These contrary effects lead to a maximum average deflection times. In Fig. 2(b), we see that the packet loss rate $\eta$ decreases until becoming zero when the node delivery coefficient $C$ increases, the reason of which is obvious. In Fig. 2(c), we see that the average arrival time $T_a$ increases with the node delivery coefficient $C$ first,  and then decreases until becoming 6,  which is the average transmission path length of the efficient routing protocol. There is also a maximum average arrival time. The increase of average deflection times results in the increase of the average arrival time. However, when the node delivery capability is large enough, the number of deflected packets decreases until becoming zero, which results in the decrease of the average arrival time.
 \begin{figure*}[!t]
 \centering
 \includegraphics[width=5in,height=1.8in]{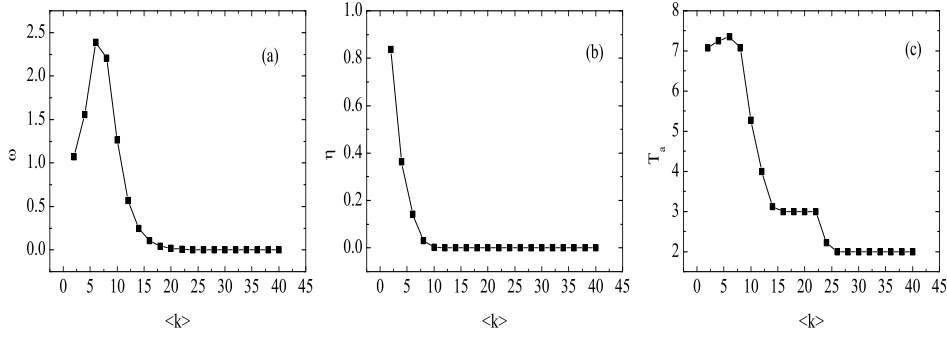}
 \caption{Average node degree $\langle k \rangle$ vs. (a) average deflection times $\omega$, (b) packet loss rate $\eta$, and (c) average arrival time $T_a$. The other parameters are: network size $N=1000$,  power-law parameter $\gamma=3$, packet generation rate  $\rho=2$, node delivery coefficient $C=1$, and routing control parameter $\alpha=1$.   Each data point is the average of 100 independent runs. }
 \end{figure*}
 \begin{figure*}[!t]
 \centering
 \includegraphics[width=5in,height=1.8in]{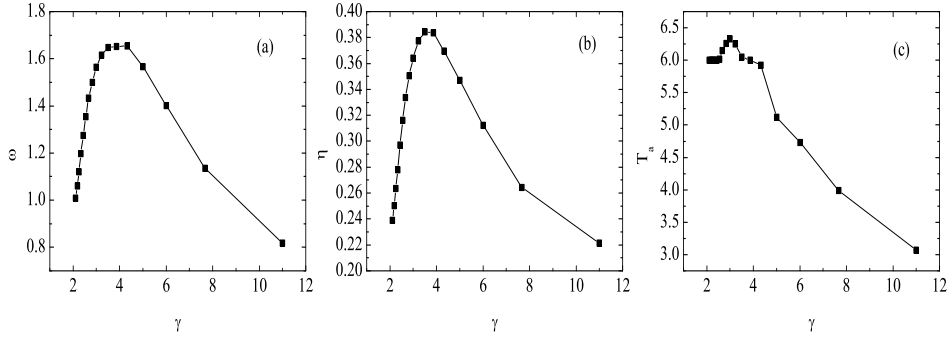}
 \caption{Power-law parameter  $\gamma$ vs. (a) average deflection times $\omega$, (b) packet loss rate $\eta$, and (c) average arrival time $T_a$. The other parameters are: network size $N=1000$,  average node degree $\langle k \rangle=4$, packet generation rate  $\rho=2$, node delivery coefficient $C=1$, and routing control parameter $\alpha=1$.   Each data point is the average of 100 independent runs. }
 \end{figure*}

 Next, we study the influence of network topological structures on the transmission performances. First, we vary the average node degree to see the change of the relevant transmission performances. In the experiments, we set network size $N=1000$, power-law parameter $\gamma=3$, packet generation rate $\rho=2$, node delivery coefficient $C=1$, and routing control parameter $\alpha=1$. The simulation results are shown in Fig. 3, where each data point is the average of 100 independent runs. We see that the variation trend in the curves of  the average deflection times $\omega$, the packet loss rate $\eta$, and the average arrival time $T_a$  is similar to that in  Fig. 2. In fact, the increase of average node degree also increases the node delivery capability according to our definition of node degree capability. Thus, the explanation of the results in Fig. 2 also applies to the results in Fig. 3.  On the other hand, the average transmission path length decreases with the increase of the average node degree, which  facilitates the data transmission. This is why the curves of Fig. 3 change more intensely than Fig. 2.  Then, we vary the power-law parameter $\gamma$ to change the degree of network heterogeneity, and study how the transmission performances change. The relevant parameters are as follows: network size $N=1000$, average node degree $\langle k \rangle=4$, packet generation rate $\rho=2$, node delivery coefficient $C=1$, and routing control parameter $\alpha=1$. The simulation results are shown in Fig. 4, each data point is the average of 100 independent runs. In Fig. 4(a), we see that the deflection times $\omega$ increases with power-law parameter $\gamma$ first, and then decreases slightly. The larger the $\gamma$, the more homogeneous the node degree distribution is. Homogeneous node degree distribution generally makes the neighbor nodes more capable to deflect the packets.  Moreover, the average transmission path becomes longer with the increase of the homogeneity of node degree distribution, which means the probability of being deflected increases. These effects lead to the increase of the average deflection times. On the other hand, when the node degree distribution becomes even, the packets will be distributed more even and the forwarding competition becomes less, and this is why the average deflection times decrease. In Fig. 4(b), we see that the packet loss rate $\eta$ goes up with the increase of power-law parameter $\gamma$, and then goes down. The  average transmission path becomes longer when the node degree distribution becomes more even, which results in the increase of packet loss rate.   On the other side, the forwarding competition becomes less when the node degree distribution becomes more evenly, which results in less packet loss. In Fig. 4(c), we see that the average arrival time $T_a$ increases first with the power-law parameter $\gamma$, and then decreases.  The increase of both the average path length and the average deflection times results in the increase of the average arrival time. Accordingly, the decrease of the average deflection times results in the decrease  of the average arrival time. Next, we vary the number of nodes while fixing the other parameters to study the change of the transmission performances. In the experiments, we set average node degree $\langle k \rangle=4$, power-law parameter $\gamma=3$, packet generation rate $\rho=1$, packet delivery coefficient $C=2$, and the routing control parameter $\alpha=1$. The simulation results are given in Fig. 5, where each data point is the average of 100 independent runs. From Fig. 5, we see that all the three performances $\omega$, $\eta$, and $T_a$  increase with the number of nodes $N$. More nodes result in more packets inserted into the network at each time step while the node delivery capability is fixed. Moreover, the average transmission path length increases with the number of nodes. All these reasons lead to the deterioration of the transmission performances.
\begin{figure*}[!t]
\centering
\includegraphics[width=5in,height=1.8in]{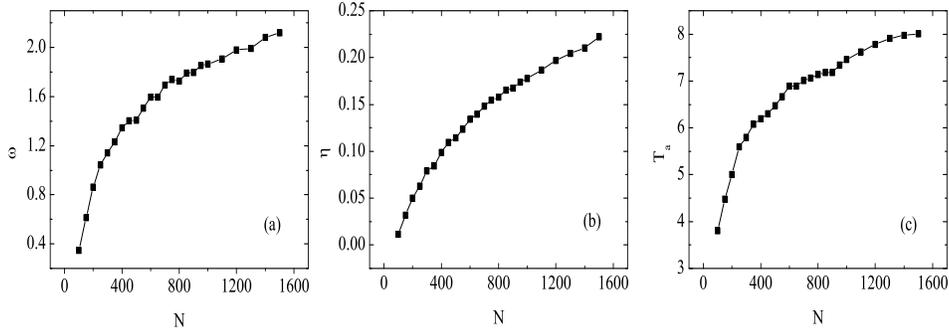}
\caption{Network size $N$ vs. (a) average deflection times $\omega$, (b) packet loss rate $\eta$, and (c) average arrival time $T_a$. The other parameters are:   average node degree $\langle k \rangle=4$, power-law parameter $\gamma=3$, packet generation rate  $\rho=1$, node delivery coefficient $C=2$, and routing control parameter $\alpha=1$.   Each data point is the average of 100 independent runs. }
\end{figure*}
\begin{figure*}[!t]
\centering
\includegraphics[width=5in,height=1.8in]{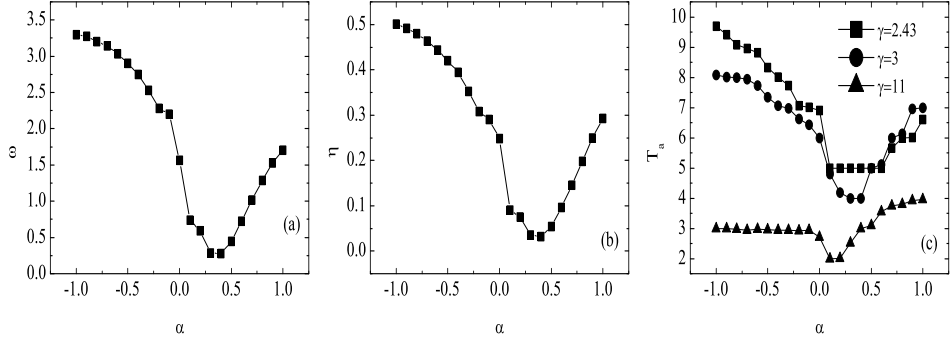}
\caption{Routing control parameter $\alpha$ vs. (a) average deflection times $\omega$, (b) packet loss rate $\eta$, and (c) average arrival time $T_a$. The common parameters of the three sub-figures are:   network size $N=1000$, average node degree $\langle k \rangle=4$,  packet generation rate  $\rho=1$, and node delivery coefficient $C=1$.  For (a) and (b), the power-law parameter is $\gamma=3$, while for (c), $\gamma$ is marked in the legend.  Each data point is the average of 100 independent runs. }
\end{figure*}

Finally, we study how the routing strategy affects the transmission performances. In the experiments, we vary the routing control parameter $\alpha$ while fixing the other parameters, which are network size $N=1000$, average node degree $\langle k \rangle=4$, power-law parameter $\gamma=3$, packet generation rate $\rho=1$, node delivery coefficient $C=1$. The simulation results are given in Fig. 6, where each data point is the average of 100 independent runs. From Fig. 6, we see that the three transmission performances $\omega$, $\eta$ and $T_a$ have the same varying trend. They first goes down with the increase of the routing control parameter $\alpha$, and then goes up. There is an optimal routing control parameter around 0.4, which corresponds to the minimum $\omega$, $\eta$ and $T_a$.  $\alpha \approx 0.4$ means the packets are prone to pass through small-degree nodes in order to achieve good transmission performances. Also, the shortest path routing protocol (corresponding to $\alpha=0$ in the efficient routing protocol), commonly used in the bufferless communication networks, is not the optimal choice.
 Note that the optimal routing control parameter is dependent on the specific experimental conditions. As shown in Fig. 6(c), the optimal routing control parameter is  different for different values of  power-law parameter.

\section{Conclusion}
In summary, we study the data transmission in bufferless communication networks from the perspective of network science. Instead of network capacity well discussed in literature, we focus on the transmission performances such as the deflection times and packet loss rate, since these metrics are the main concerns for the bufferless communication networks. We apply the Price model and the efficient routing protocol to the context of bufferless data transmission.  We propose a delivery queue based deflection mechanism to deal with the packet forwarding competition. We find that the effects of factors, such as the packet generation rate, node delivery capability,  average node degree, and degree distribution, on the transmission performances  are  in general non-monotonic. As for the efficient routing protocol, there is optimal control parameter dependent on the network conditions, leading to the optimal transmission performances. Our work is a preliminary effort to apply theories and tools from network science to bufferless communication networks. We believe it will help to  build  the network architecture and network routing protocol  in the burfferless communication systems.


%



\section*{Acknowledgment}

This work was  supported by the National Natural Science Foundation of China (Grant No. 61304154).

\ifCLASSOPTIONcaptionsoff
  \newpage
\fi

\end{document}